# Intrinsic vs Extrinsic Spin Currents. Old Ideas in a New Light.

Alexander Khaetskii

*Institute of Microelectronics Technology, Russian Academy of Sciences, 142432, Chernogolovka, Moscow District, Russia*

**Abstract.** We have described the electron spin dynamics in the presence of Rashba spin-orbit interaction and disorder using the spin-density matrix method. We showed that in the Born approximation in the scattering amplitude the spin current is zero for an arbitrary ratio of the spin-orbit splitting and the scattering rate and for an arbitrary disorder potential. We also describe some magnetotransport phenomena such as negative magnetoresistance and a negative charge Hall effect which occur in the presence of spin-orbit coupling.

## INTRODUCTION

Spin-orbit coupling brings about a number of interesting effects, one of which is generation of a spin flux in the plane perpendicular to the charge current. We use the term spin flux or spin current, which is the same. This phenomenon occurs in the paramagnetic system and has been known for quite a long time [1]. It is a consequence of the fact that in the presence of spin-orbit coupling the scattering by impurities has an asymmetric character (the Mott effect) [2]. Spins with up-orientation are scattered preferably to the right and with down-orientation - to the left. This phenomenon exists only beyond the Born approximation in the scattering amplitude and leads to an accumulation of the spin density near the sample surface [1]. Mutual transformation of the current and spin fluxes leads also to the renormalization of the electrical conductivity of the system [3].

It has been recently claimed [4,5] that an analogous phenomenon can exist even without scattering by impurities, i.e. in the ballistic regime, the corresponding contribution being called intrinsic. In particular, in the case of the 2D electron system described by the Rashba Hamiltonian the universal value $e/8\pi\hbar$ for the spin current was derived. Later several papers appeared where the effect of scattering by impurities was taken into account with a range of totally different results. This was done within the diagrammatic Kubo approach.

We solve this problem using the well known method of a spin-density matrix [3]. We show by exact calculations, keeping all the components of the spin density matrix, that in the case of the Rashba Hamiltonian the intrinsic spin current does not exist. In the Born approximation (when the scattering amplitude has additional symmetry properties) the spin current is found to be zero for an arbitrary value of $\Delta\tau_{tr}$, where $\Delta = \alpha p_F$ is the spin splitting of the electron spectrum and $\tau_{tr}$ the transport scattering time ($\alpha$ is the spin-orbit coupling constant and $p_F$ Fermi momentum). It should be noted that calculations for other Hamiltonians (for example 3D and 2D holes) [6, 7] give a nonzero result for the spin current already in the Born approximation. For example, in the case of 2D holes in the limit $\Delta \gg 1/\tau_{tr}$ and for the short range impurity potential we obtain for the spin conductivity $3e/8\pi\hbar$. [7] However, in the same ballistic limit but for a smooth impurity potential we obtain twice as large number [7], i.e. the result explicitely depends on the disorder properties and does not have the universal value.

## Hamiltonian, Spin Current. Kinetic Equation

The Hamiltonian of the problem has the form:

$$\hat{H} = \frac{p^2}{2m} + \frac{1}{2}\alpha\vec{\sigma}\vec{\Omega}(\vec{p}), \quad \vec{\Omega}(\vec{p}) = [\vec{n}\times\vec{p}] \quad (1)$$

where $\vec{n}$ is the unit vector normal to the 2D plane (z-axis). We will calculate the $q_{yz}$ component of the spin current, where we use usual hermitian definition of this quantity:

$$q_{yz} = Tr\int \frac{d^2p}{(2\pi)^2}\hat{f}(\mathbf{p})\frac{1}{2}(\hat{S}_z\hat{V}_y + \hat{V}_y\hat{S}_z) \quad (2)$$

Here $\hat{f}(\vec{p})$ is the spin density matrix, $\hat{V}_y$ the y-component of the velocity operator. The general expression for the quantum kinetic equation in the case of the spin-orbit interaction, when the Hamiltonian and the Wigner distribution function are matrices over the spin indexes, was derived in [3]. In our case when we deal only with the electric field **E** which is constant in space this equation is simple and reads

$$\frac{\partial\hat{f}(\mathbf{p})}{\partial t} + e\mathbf{E}\frac{\partial\hat{f}(\mathbf{p})}{\partial\mathbf{p}} + \frac{i}{\hbar}[\hat{H}(p), \hat{f}(\mathbf{p})] = St\{\hat{f}(\mathbf{p})\} \quad (3)$$

It differs from the common classical kinetic equation by the last term on the left hand side which is the commutator of the Hamiltonian with the spin density matrix. This commutator is due to the spin-orbit splitting of the electron spectrum and leads to the precession of the spin around the effective momentum-dependent magnetic field. The collision term was derived in many papers, and in the helicity basis has the form

$$St(\hat{f}(\mathbf{p}))_{MM'} \propto \int d^2p_1 \sum_{M_1 M_1'} K_{M_1 M_1'}^{MM'} f_{M_1 M_1'}(\mathbf{p}_1) \quad (4)$$

where the kernel in the Born approximation in the scattering amplitude is ($\theta$ is the scattering angle):

$$K_{M_1 M_1'}^{MM'}(\theta) \propto \delta(\varepsilon_{M_1}(p_1) - \varepsilon_M(p))F_{M_1\mathbf{P}_1}^{M\mathbf{P}}(F_{M_1'\mathbf{P}_1}^{M'\mathbf{P}})^* \quad (5)$$

here $\varepsilon_M(p)$ are the eigenvalues, M= $\pm 1/2$ the helicity values. The Born scattering amplitude is given by:

$$F_{M_1\mathbf{P}_1}^{M\mathbf{P}} \propto D_{MM_1}^{(1/2)}(\theta)U(\mathbf{p}-\mathbf{p}_1), \quad F_{M_1\mathbf{P}_1}^{M\mathbf{P}} = (F_{M\mathbf{P}}^{M_1\mathbf{P}_1})^* \quad (6)$$

$U(\mathbf{p}-\mathbf{p}_1)$ is the Fourier component of the impurity potential. The additional symmetry property indicated in Eq.6 exists only in the Born approximation [2]. It means the equality of the scattering amplitudes for the direct and reverse processes, which are obtained by interchanging the initial and final momenta without changing their signs.

*Smooth scattering potential.*

First consider the mathematically simple case of a smooth scattering potential when the interband transitions, i.e. the ones between the energy bands corresponding to opposite helicity values, are suppressed. This case is realized at $m\alpha R/\hbar \gg 1$, where R is the radius of impurity. Then from Eqs.(3,4) we obtain:

$$eE\frac{\partial f_+^{(0)}}{\partial p} = \frac{2ap}{V_+}f_{++} + \frac{bp}{V_+}(f_{+-} - f_{-+}),$$

$$eE\frac{\partial f_-^{(0)}}{\partial p} = \frac{2ap}{V_-}f_{--} - \frac{bp}{V_-}(f_{+-} - f_{-+}), \quad (7)$$

$$\frac{ieE}{2p}(f_+^{(0)} - f_-^{(0)}) + \frac{i}{\hbar}(\varepsilon_+ - \varepsilon_-)f_{+-} =$$
$$c(\frac{p}{V_+}f_{++} - \frac{p}{V_-}f_{--}) + pd(\frac{1}{V_+} + \frac{1}{V_-})f_{+-}, \quad (8)$$

$$\frac{-ieE}{2p}(f_+^{(0)} - f_-^{(0)}) - \frac{i}{\hbar}(\varepsilon_+ - \varepsilon_-)f_{-+} =$$
$$-c(\frac{p}{V_+}f_{++} - \frac{p}{V_-}f_{--}) + pd(\frac{1}{V_+} + \frac{1}{V_-})f_{-+} \quad (9)$$

where d=a, b=-c, c=-ia,

$$a = -\frac{1}{2}\int\frac{d\theta}{2\pi}W(\theta)\sin^2\theta \quad (10)$$

quantity $-2ap/V_+$ is equal to the inverse transport scattering time $1/\tau_{tr}$, $W(\theta) = N|U(p-p_1)|^2/2\hbar^3$, $f_+^0(p), f_-^0(p)$ are the equilibrium Fermi functions which correspond to the helicity +/-, $V_{+/-}(p) = p/m \pm \alpha/2$ are the velocity values for a given momentum p for +/- bands. The quantities entering Eqs.(7-9) have the following relations to the average spin components:

$$<S_z> \propto (f_{+-} + f_{-+}), <\mathbf{S}\cdot\mathbf{p}> \propto (f_{+-} - f_{-+}), \quad (11)$$
$$<\mathbf{S}\cdot\mathbf{\Omega}> \propto (f_{++} - f_{--}).$$

From the above equations for the quantity of interest ($<S_z>$) we find:

$$eE(\frac{\partial f_+^0}{\partial p} - \frac{\partial f_-^0}{\partial p}) + \frac{eE}{p}(f_+^0 - f_-^0) =$$

$$= -\frac{1}{\hbar}(\varepsilon_+ - \varepsilon_-)(f_{+-} + f_{-+}) \qquad (12)$$

This equation is exact for an arbitrary value of $\Delta\tau_{tr}$. After integration in Eq.(2) we obtain $q_{yz} = 0$, i.e. the spin current is zero. Note that Eq.(12) has a clear physical meaning. The second term on the left hand side was taken into account before [4] and describes the appearance in the electric field of the z-component of the spin due to the angular dependence of the wave functions. Exactly this term gives the contribution $-e/8\pi\hbar$ after integration in Eq.(2). However, the first term in Eq.(12) describes the change in the distribution functions due to the acceleration along the electric field and cancels exactly the contribution of the second term after integration in Eq.(2). Note that the result of [4] can be obtained if Eq.(9) is subtracted from Eq.(8) and the right hand side (collision term) is neglected altogether. This is exactly what was done by the authors of Ref. [4] since they solved the equations of motion for the spin totally ignoring the collision term. The solution of collisionless equations gives a wrong result because the neglected terms give the contribution of the same order (for an arbitrary large $\tau_{tr}$) as the term which was kept in Ref.[4].

Eq.(12) can also be obtained from Eq.(3) if one calculates the mean value of

$$<S_i> = \frac{1}{2}\int d^2p\, Tr\{\hat{\sigma}_i \hat{f}(\mathbf{p})\}, i = y \qquad (13)$$

Then the contribution from the second term of Eq.(3) gives the left hand side of Eq.(12) and from the third one - the right hand side of Eq.(12). The contribution from the collision term is zero since
in the Born approximation there is no spin relaxation due to rotation of the spin during the collision event itself (for an arbitrary impurity potential). Finally, Eq.(12) is equivalent to calculating $q_{yz}$ through the following formula [8]:

$$q_{yz} \propto eE \int d^2p\, \frac{\partial S_y^{(0)}(\vec{p})}{\partial p_x} = 0 \qquad (14)$$

where $S_y^{(0)}(\vec{p})$ is the equilibrium value of the y-spin component for the given momentum.

## Magnetotransport Phenomena due to Spin-orbit Coupling

As it is known [1, 3], due to spin-orbit coupling there exists mutual transformation of particle and spin fluxes. This phenomenon occurs beyond the Born approximation in the scattering amplitude. This transformation of the particle and spin fluxes leads to the renormalization of the momentum relaxation time so that the conductivity gets reduced [3]. In this section we consider an influence of spin-orbit interaction on the magnetoresistance and the charge Hall effect in semiconductors. The magnetic field is assumed to be classical (i.e. no quantization of the electron spectrum is taken into account but the parameter $\omega_c\tau_{tr}$ can be arbitrary). We consider only one type of carriers-electrons which are strongly degenerate (i.e. temperature T<<$E_F$). In semiconductors with a simple conduction band (like InSb, GaAs) magnetoresistance is zero when spin-orbit coupling is neglected, i.e. $\sigma(H) = \sigma(0)$. The Hall constant does not depend on magnetic field and has a simple form R(H)=1/nec, where n is the electron concentration. The external magnetic field influences not only the momentum of the particles but also their spin flux, namely it destroys it. As a result, the magnetoresistance in 3D case (both longitudinal and transverse) becomes negative [9]. This result can be easily understood. Since in the absence of magnetic field the spin flux reduces the conductivity [3], it is quite natural, its destruction by the external magnetic field leads to the increase of the conductivity (i.e. negative magnetoresistance), see Fig.1. Hall constant becomes magnetic field dependent and has a simple form indicated above only in the classically strong magnetic field ($\omega_c\tau_{tr}$ >1), but in the classically weak magnetic field it contains the Hall-factor which depends on the spin-orbit coupling constant.

We need to write down the coupled system of the equations determining the time evolution of the mean momentum $\bar{p}$ and mean spin flux $\bar{q}_{\alpha\beta}(q_{\alpha\beta} = p_\alpha S_\beta)$ of the particles with fixed energy. The bar means averaging with the density matrix over the momentum direction and the trace over the spin variables. These equations are:

$$\frac{dp_i}{dt} = eE_i - \omega_c[\mathbf{p}\times\mathbf{h}]_i - \frac{p_i}{\tau_{tr}} + \beta\varepsilon_{ikl}q_{kl} \qquad (15)$$

$$\frac{dq_{\alpha\beta}}{dt} = -\omega_c\varepsilon_{\alpha kl}h_l q_{k\beta} - \Omega_s\varepsilon_{\beta kl}h_l q_{\alpha k} - \frac{q_{\alpha\beta}}{\tau_{tr}} - \frac{\beta}{4}\varepsilon_{\alpha\beta\gamma}p_\gamma$$

In Eqs.(15) **E,H** are electric and magnetic fields, $\omega_c, \Omega_s$ - cyclotron and spin precession frequencies, **H** || z. The last term in the first equation (15) describes

the occurrence the particle flux perpendicular to the spin flux upon scattering. Analogously, the last term in second equation (15) describes the occurrence of the spin flux. Thus, $\beta$ is the coupling between the particle flux and the spin flux and is proportional to the spin-orbit coupling strength, $\beta \propto \tilde{\alpha} e^2 / \hbar v_F$, where $\tilde{\alpha}$ is an effective s-o coupling and $e^2 / \hbar v_F$ the Born parameter for the scattering by the charged impurities. In the case of weak spin-orbit coupling considered here $\beta \tau_{tr} \ll 1$. The two first terms in the right hand side of second equation (15) describe the influence of the magnetic field on the spin flux due to the rotation of the momentum and the spin correspondingly. The third term describes the spin flux relaxation upon scattering. The result for the longitudinal conductivity $\sigma_l(H)$ in the 3D case reads:

$$\sigma_l(H) = \frac{ne^2 \tau_{tr}}{m}[1 - \frac{1}{2}\frac{\beta^2 \tau_{tr}^2}{1+(\Omega_s - \omega_c)^2 \tau_{tr}^2}], \quad (16)$$

As follows from Eq.(16), the conductivity increases with magnetic field and saturates in classically strong magnetic fields, see Fig.1. For the transverse magnetoresistance we obtain:

$$\sigma_t(H) = \sigma_{xx} + \frac{\sigma^2_{xy}}{\sigma_{xx}} = \quad (17)$$

$$\frac{ne^2 \tau_{tr}}{m}[1 - \frac{\beta^2 \tau_{tr}^2}{4}(\frac{1}{1+\omega_c^2 \tau_{tr}^2} + \frac{1}{1+\Omega_s^2 \tau_{tr}^2})]$$

The transverse conductivity also increases with magnetic field because of the destruction of the spin fluxes by the magnetic field. The Hall constant takes the form:

$$R = (\frac{\sigma_{xy}}{\sigma_{xx}^2 + \sigma_{xy}^2})\frac{1}{H} =$$

$$\frac{1}{nec}[1 - \frac{\beta^2 \tau_{tr}^2}{4}(\frac{1}{1+\omega_c^2 \tau_{tr}^2} + \frac{\Omega_s}{\omega_c}\frac{1}{1+\Omega_s^2 \tau_{tr}^2})]$$

(18)

Thus, the Hall constant has a simple form 1/nec only in the limit of a classically strong magnetic field but in a weak field it contains a Hall factor. The reason for the change of the Hall constant can be easily understood. Since the spin flux is perpendicular to the current direction (x-axis), in the absence of magnetic field only the components $q_{yz}, q_{zy}$ of the spin flux tensor are not equal to zero. The magnetic field rotating the particle momentum moves the spin-down electrons in the positive x-direction and spin-up electrons- in the negative x-direction. It means that the $q_{xz}$ component becomes nonzero. One can also say that the magnetic field rotates the plane where the spin fluxes flow. In accordance with Eq.(15), the occurrence of the $q_{xz}$ component means that in the y-direction the charge current appears additional to that due to the Lorentz force. This means in turn that the Hall constant changes.

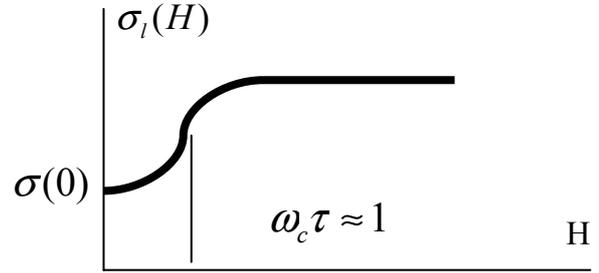

**FIGURE 1.** Dependence of the inverse longitudinal resistance on the external classical magnetic field.

## ACKNOWLEDGMENTS

I am grateful to M.I. D'yakonov, L. Glazman and A.H. MacDonald for fruitful discussions.